1

# SOME PROPERTIES OF $YBa_mCu_{1+m}O_y$ ($m$ = 2,3,4,5) SUPERCONDUCTORS

PIYAMAS CHAINOK
*Prasarnmit Physics Research Unit, Department of Physics, Faculty of Science,*
*Srinakharinwirot University, Bangkok 10110, Thailand.* AND
*Department of Science and Mathematics, Faculty of Science and Technology,*
*Pathumwan Institute of Technology, Bangkok 10330, Thailand.*
*bpiyamas@hotmail.com*

THANARAT KHUNTAK
*Prasarnmit Physics Research Unit, Department of Physics, Faculty of Science,*
*Srinakharinwirot University, Bangkok 10110, Thailand.*
*tkiambas@gmail.com*

SUPPHADATE SUJINNAPRAM
*Department of Physics, Faculty of Liberal Arts and Science,*
*Kasetsart University, Kamphaeng Saen Campus, Nakhon Pathom 73140, Thailand.* AND
*Thailand Center of Excellence in Physics(ThEP), Si Ayutthaya Road, Bangkok 10400, Thailand.*
*supphadate.s@ku.ac.th*

SOMPORN TIYASRI
*Department of Chemistry, Faculty of Science,*
*Srinakharinwirot University, Bangkok 10110, Thailand.*
*somporn.swu@gmail.com*

WIRAT WONGPHAKDEE
*Department of Chemistry, Faculty of Science,*
*Srinakharinwirot University, Bangkok 10110, Thailand.*
*wirat@swu.ac.th*

THITIPONG KRUAEHONG
*Department of Physics, Faculty of Science and Technology,*
*Suratthani Rajabhat University, Surat Thani 84100, Thailand.*
*kruaehong@hotmail.com*

TUNYANOP NILKAMJON
*Prasarnmit Physics Research Unit, Department of Physics, Faculty of Science,*
*Srinakharinwirot University, Bangkok 10110, Thailand.* AND
*Thailand Center of Excellence in Physics(ThEP), Si Ayutthaya Road, Bangkok 10400, Thailand.*
*swu009@gmail.com*

SERMSUK RATRENG
*Prasarnmit Physics Research Unit, Department of Physics, Faculty of Science,*
*Srinakharinwirot University, Bangkok 10110, Thailand.* AND
*Thailand Center of Excellence in Physics(ThEP), Si Ayutthaya Road, Bangkok 10400, Thailand.*
*ser_rat@hotmail.com*

PONGKAEW UDOMSAMUTHIRUN
*Prasarnmit Physics Research Unit, Department of Physics, Faculty of Science,*
*Srinakharinwirot University, Bangkok 10110, Thailand.* AND
*Thailand Center of Excellence in Physics(ThEP), Si Ayutthaya Road, Bangkok 10400, Thailand.*
*udomsamut55@yahoo.com*

abstract

We synthesized the $YBa_mCu_{1+m}O_y$ superconductors; $m$ = 2,3,4,5 that were Y123 ($YBa_2Cu_3O_{7-x}$), Y134 ($YBa_3Cu_4O_{9-x}$), Y145 ($YBa_4Cu_5O_{11-x}$), Y156 ($YBa_5Cu_6O_{13-x}$), by solid state reaction with the $Y_2O_3$, $BaCO_3$ and $CuO$ as the beginning materials. The calcination temperature was 950 °C and varied the sintering temperature to be 950 °C and 980 °C. The resistivity measurement by four-point-probe technique showed that the $T_c^{onset}$ of Y123, Y134, Y145, Y156 were at 97 K, 93 K, 91K, 85 K, respectively. The XRD and Rietveld full-profile analysis method were used and found that the crystal structure was in the orthorhombic with Pmmm space group with the ratio $c/a$ were 3.0, 4.0, 5.0 and 6.0 for Y123, Y134, Y145 and Y156, respectively. The oxygen content was characterized by Iodometric titration. The ($Cu^{3+}/Cu^{2+}$ and Oxygen content) were (0.28, 6.83), (0.19, 8.81), (0.13, 10.79), (0.16, 12.92) of Y123, Y134, Y145, Y156 respectively. We also found that the increasing of sintering temperature has reduced the oxygen content and the critical temperature of all samples.

## 1. Introduction

The cuprate superconductor has become one of the most interested superconductors due to its many potential applications such as magnetic levitation transportation, microwave devices, and power transmission tape. The $YBa_2Cu_3O_{7-x}$ (Y123) superconductor is first superconductor having critical temperature ($T_c$) more than the boiling point of liquid nitrogen. The first Y123 superconductor had been synthesized by Chu and co-workers[1] since 1986 with the critical temperature at 92 K. In the orthorhombic perovskites crystal structure of Y123, there are two different $Cu$ sites: $CuO$ chains and $CuO_2$ planes, i.e., $Cu$ (1) site in $CuO$ chains and $Cu$ (2) in $CuO_2$ planes. The $Cu$ (1) atom coordination is square planar for the $CuO$ chains and $Cu$ (2) atom coordination is square pyramidal in the $CuO_2$ planes. The $CuO_2$ planes play an important role of superconductivity, whereas $CuO$ chains are non-superconducting[2] and the crystal structure occurs an alternating a superconducting $CuO_2$ plane and a blocking layer along the c-axis direction. The hopping interaction between these leads to the suppression of $T_c$[3, 4]. The substitution at the $Cu$ (1) site can cause an increase in the oxygen content in the $CuO_2$ plane that can induce to occur the orthorhombic to tetragonal (O–T) transition. The oxygen in the basal plane acts as a charge reservoir introducing holes into the $CuO_2$ plane. If the substitution takes place on the $Cu$ (2) site, there is no observed structural O–T transition that the structure remains orthorhombic[5]. The substitution of $Cu$ by non-magnetic ions such as $Zn$, $Ca$, and $Ga$ or magnetic ions ($Cr$, $Co$, $Fe$) in the cuprate superconductors is a useful tool for probing superconductivity parameters[6]. Dong Han Ha et al.[7] studied the effects of cation substitution, $Sr$ and $Ca$, on the oxygen loss in $YBaCuO$ superconductors. They found that oxygen is removed more easily with increasing Ca concentration which may be due to the displacement of Ba ions from the $Cu$-$O$ chain towards the $CuO_2$ plane. They found the oxygen content at y = 6.96 for $YBa_2Cu_3O_y$ with $T_c \approx 92$K. Sahoo, and Behera[8] studied the effect of transition metal Cr substitution on structural, microstructure, and the electrical resistivity of $YBa_2Cu_3O_y$ superconductor.

During the past twenty years, the researchers have been carried out on the $YBaCuO$-family compounds like Y123, $YBa_2Cu_4O_8$ (Y124), and $Y_2Ba_4Cu_7O_{15}$ (Y247). They found that Y124 and Y247 became superconductor at 80 K [9] and 40K [10], respectively. The Y247 exhibits a superconducting transition with $T_c$ ranging from 30 to 95 K, depending on the oxygen content[11, 12]. In 2009, Aliabadi, Farshchi and Akhavan et al.[13] and Tavana[14] synthesized Y358 ($Y_3Ba_5Cu_8O_y$) superconductor by solid state reaction that become superconducting above 100 K with the lattice parameters a = 3.888 Å, b = 3.823 Å, c = 31.013 Å. The Y123 has two $CuO_2$ planes and one $CuO$ chain. The Y124 has one $CuO$ double chain. The Y247 has one $CuO_2$ planes and one $CuO$ chain, and one double chain .The Y358 has crystal structure similar to Y123 with five $CuO_2$ planes and three $CuO$ chains. The increasing in the number of $CuO_2$ planes and $CuO$ chain have important effect on the $T_c$ of $YBaCuO$ superconductor. In 2010, Udomsamuthirun et al.[15] found the new $YBaCuO$ superconductors; Y5-6-11, Y7-9-16, Y5-8-13, Y7-11-18, Y1-5-6, Y3-8-11 and Y13-20-33. These superconductors were synthesized by using the assumption that the number of $CuO_2$ planes and $CuO$ chains have related to the number of $Ba$-atom and $Y$-atom and the number of $Ba$-atom plus $Y$-atom are equal to the number of $Cu$-atom. Topal and Akdogan[16] synthesized and characterized three new $YBaCuO$ superconductors i.e., $Y_2Ba_3Cu_{5.2}O_y$ (Y2352), $Y_2Ba_5Cu_9O_y$ (Y259), and $Y_1Ba_4Cu_5O_y$ (Y145). The $T_c^{onset}$ was determined to be 98 K, 98 K, and 97.3 K for Y-2352, Y-145, and Y-259, respectively. The x-ray analysis show that they have a similar crystalline structure as Y-123 phases. Chainok et al.[17] synthesized and characterized the physical properties of Y123 and $Y_1Ba_4Cu_5O_y$ (Y145) superconductors by solid state reaction reacting in the air atmosphere with sintering temperature at 950 °C and 980 °C. The crystal structure of Y145 is orthorhombic which a = 3.80 Å, b = 3.86 Å and c = 19.37 Å and the peritectic temperature at 1018 °C is found. Murakami[18] studied the melt processing of $YBaCuO$ superconductors. In the $YBaCuO$ system, the superconducting phase is produced by a peritectic reaction: $Y_2BaCuO_5$ (Y211) + L $\rightarrow$ $2YBa_2Cu_3O_x$ (Y123). If incomplete peritectic reaction is occurred, the crystals would contain fine 211 inclusions. However, the fine dispersion of Y211 inclusions is considered to have three beneficial effects: to suppress crack formation; to promote oxygen diffusion; and to provide pinning centers. So the melt-textured of $YBaCuO$ superconductor can exhibit the different properties from bulk solid state material [19]. Antal et al.[20] investigated behaviors of $YBa_2Cu_3O_7$ (Y123) powder and $YBCO$ bulk superconductor prepared by Top Seeded Melt Growth process. They also proposed that the melting process of Y123 should begin at around 982 °C for Y123 powder and 984 °C for grinded Y123 bulk.

In this paper, the superconductor having one Yttrium atom was fabricated, the $YBa_mCu_{1+m}O_x$; $m = 2,3,4,5$ superconductors which were Y123 ($YBa_2Cu_3O_{7-x}$), Y134 ($YBa_3Cu_4O_{9-x}$), Y145 ($YBa_4Cu_5O_{11-x}$), Y156 ($YBa_5Cu_6O_{13-x}$) by solid state reaction. The calcination temperature was 950 °C and the sintering temperature at 950 °C (S950) and 980 °C (S980) had been done. All samples obtained were characterized by the SEM, EDX, XRD with Rietveld full-profile analysis, DTA, Iodometric titration and the resistivity measurement.

## 2. Experimental procedure

The bulk $YBa_mCu_{1+m}O_x$; $m$ = 2,3,4,5 superconductors were synthesized by solid state reaction. The powder of $Y_2O_3$, $BaCO_3$ and $CuO$ (99.9%) were mixed and ground in stoichiometric ratios 1:2:3, 1:3:4, 1:4:5 and 1:5:6. The resulting mixture were calcined in air at 950 °C for 24 hours twice times with the intermediate grinding. The calcined powders were reground and then pressed into pellets. After that we kept the samples in two parts; sintering at 950 °C (S950) and sintering at 980 °C (S980). Finally both samples were annealing at 500 °C in air.

All samples obtained were characterized by the SEM micrograph and EDX (JEOL JSM 6400), the XRD (Bruker D8-Discover), DTA (Perkin-Elmer 7 Series Thermal Analysis System). The oxygen content was characterized by Iodometric titration. The resistivity measurements were performed with four-point-probe technique and the crystal structures were performed by Rietveld full-profile analysis method. Energy Dispersive Spectroscopy (EDS) analysis was carried out to investigate stoichiometry and chemical composition.

## 3. Results and Discussions

The surface and compositions of all samples were studied by SEM and EDX. The surfaces with inhomogeneous texture without any impurities were found. The elongated grains of samples sintered at 950 °C of Y123 and Y156 appeared to stick each other, assembled in different masses and oriented randomly in all directions throughout the micrograph. More cracks and voids between grains were found in samples sintered at 950 °C than those of 980 °C samples. The melt on the grain boundary occurred in the Y123 and Y134 sintered at 980 °C.

The temperature-dependence normalized resistivity of samples; Y123, Y134, 145 and Y156; were shown in Fig. 1. The $T_c^{onset}$ and $T_c^{offset}$ were read out from the normalized resistivity curves that the $T_c^{onset}$ was taken as the temperature at which the tangent of the resistance versus temperature curve intersects with the tangent of the part where resistance dropped abruptly and $T_c^{offset}$ was defined as the temperature at which the electrical resistance readings reached zero. The summaries of critical temperature were shown in Table 1. We found that the highest $T_c^{onset}$ was Y123, $T_c$ = 97 K and the lowest was Y156, $T_c$ = 85 K. The sintering temperature has shown effect on Y134 and Y145 that they were in the adversative manner.

Table 1. This is the critical temperature of our samples.

| Samples | $T_c^{offset}$ (K) | | $T_c^{onset}$ (K) | |
|---|---|---|---|---|
| | S950 | S980 | S950 | S980 |
| Y123 | 92 | 82 | 97 | 92 |
| Y134 | 83 | 80 | 93 | 88 |
| Y145 | 79 | 78 | 91 | 89 |
| Y156 | 80 | 81 | 85 | 85 |

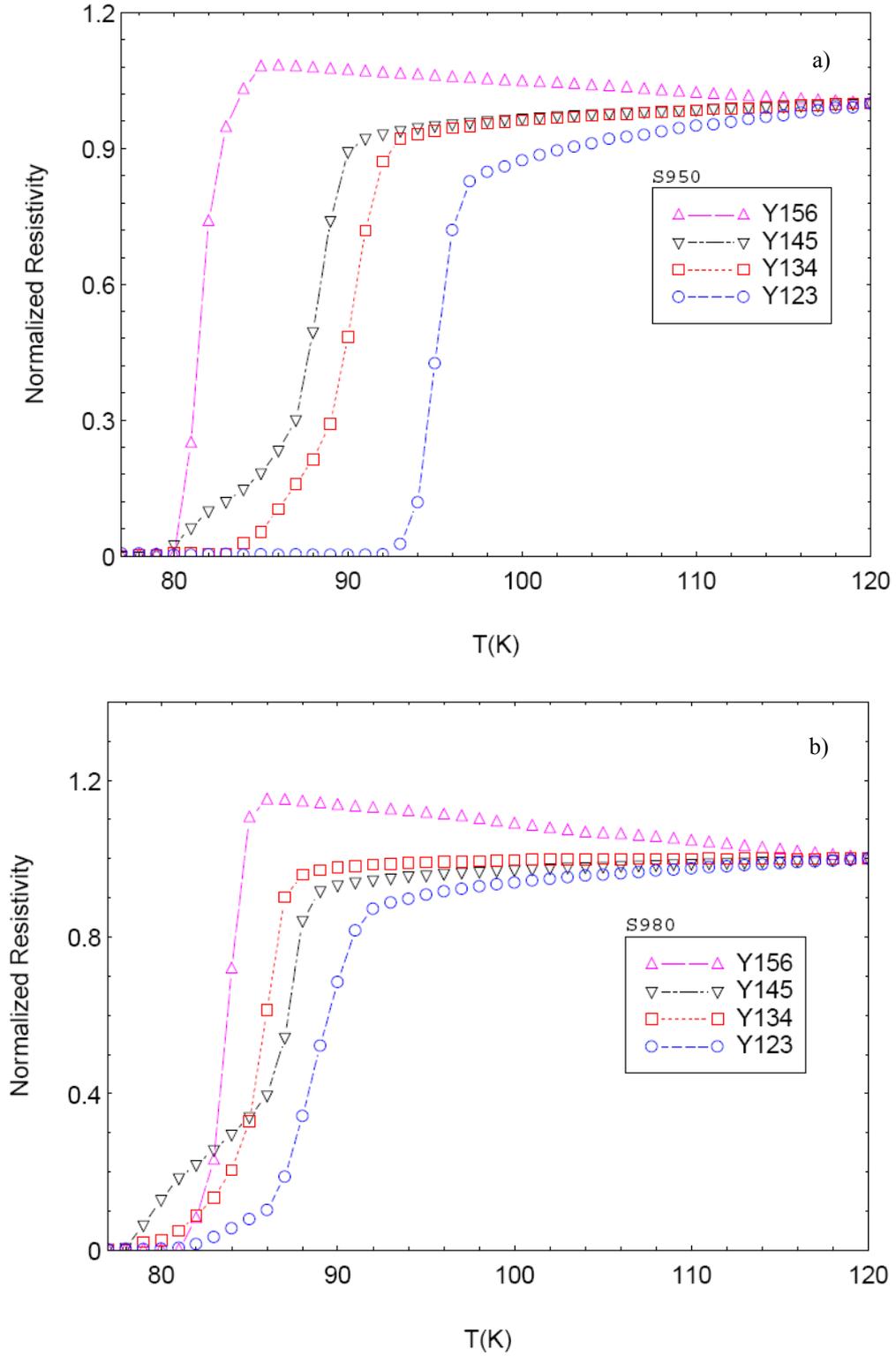

Fig. 1.  This is the normalized resistivity versus temperature of samples.
a) sintered at 950 ℃    b) sintered at 980 ℃

For the powder X-ray diffraction analysis, the pellets were reground to fine powder and then XRD analysis was carried out. The X-ray diffraction patterns taken at room temperature in $2\theta = 10° - 90°$ range were shown in Fig. 2. The characteristic peaks of the Y123, Y134, Y145 and Y156 were determined by the Full-prop software.

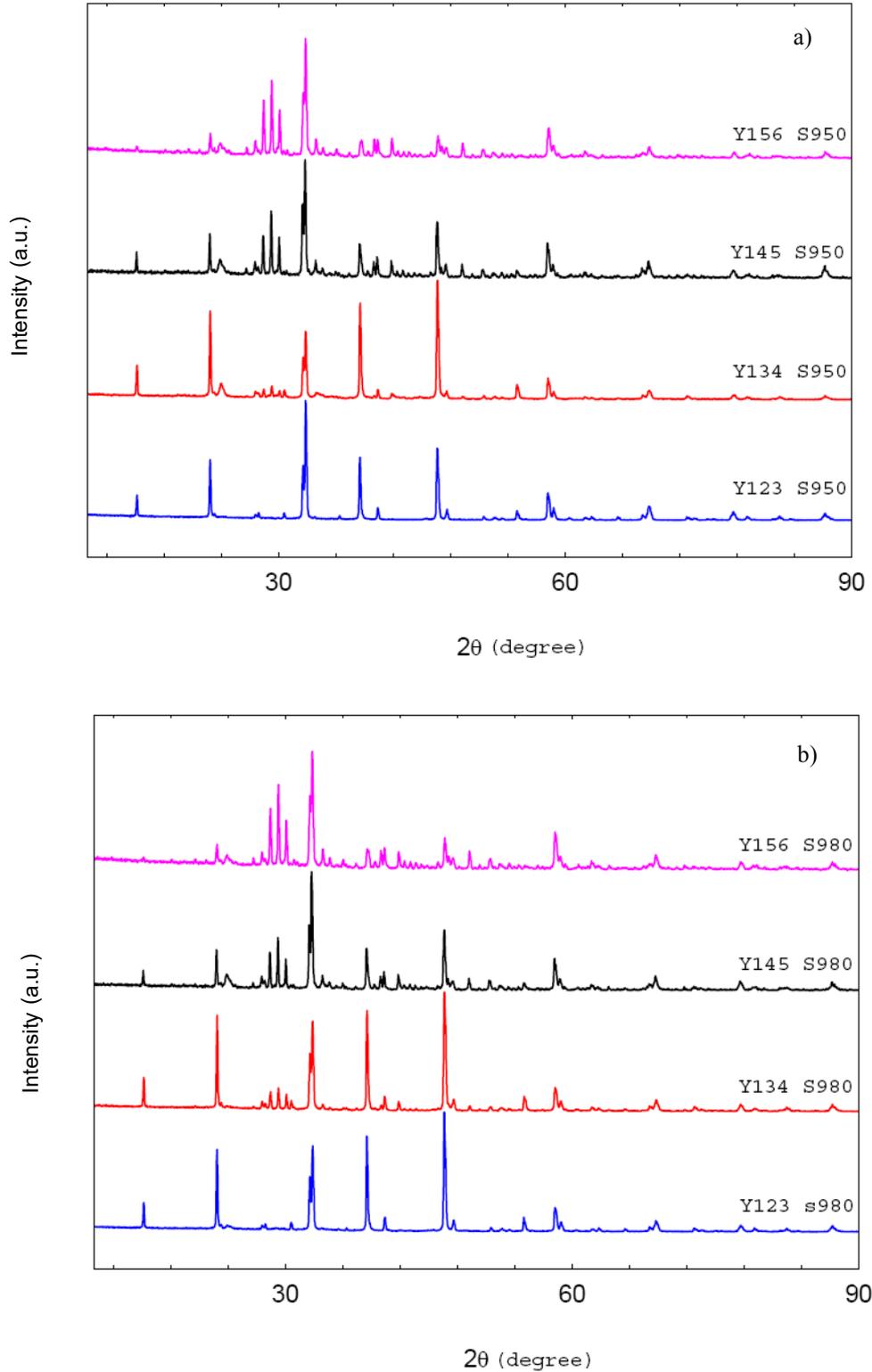

Fig. 2. The X-ray diffraction patterns of samples.
a) sintered at 950 °C    b) sintered at 980 °C

There were two compositions in our samples; superconducting compound and non-superconducting compound as shown in Table 2. The highest percentage of superconducting compound was Y123 (S950) and the lowest was Y156 (S950). The difference of sintering temperature showed little change in percentage of superconducting compound. The amount of non-superconducting compound diminish the $T_c^{offset}$ of samples that $T_c^{offset}$ of Y145 and Y156 are lower than the others.

Table 2. This is the percentage of superconducting compound and non-superconducting compound.

| Samples | | Superconducting compound (%) | Non-superconducting compound(%) |
|---|---|---|---|
| Y123 | S950 | 89.14 | 10.86 |
| | S980 | 88.25 | 11.75 |
| Y134 | S950 | 78.65 | 21.35 |
| | S980 | 78.06 | 21.94 |
| Y145 | S950 | 56.51 | 43.49 |
| | S980 | 59.61 | 40.39 |
| Y156 | S950 | 45.10 | 54.90 |
| | S980 | 46.37 | 53.63 |

The lattice parameters of superconducting compound and non-superconducting compound were shown in Table 3 and Table 4. The crystal structures of superconducting compounds were orthorhombic with Pmmm space group. There was no significant change in $a$-axis value and $b$-axis values and a change in sintering temperature has a little effect on lattice parameters. The highest $c$-axis value was of Y156. The $c/a$ was agreed with the relation between $c/a$ and number of $Cu$-atom that found by Sujinnapram et al.[21] as "$c/a$ is equal to number of $Cu$-atom minus 1". The anisotropy can be divided into two groups, less than 1.8 and higher than 2.1 that correspond to Y123, Y145 and Y134, Y156 respectively.

For Y123, the non-superconducting compound was Y211 ($Y_2BaCuO_5$) with Pbnm space group. For Y134 Y145 and Y156 the non-superconducting compounds were $BaCuO_2$ and $Ba_2Cu_3O_6$ with Im-3m and Pccm space group respectively.

Table 3. This is the lattice parameters of superconducting compound.

| Compounds | | Lattice parameter (Å) | | | Anisotropic 100(b-a)/0.5(b+a) | c/a | $\chi^2$ |
|---|---|---|---|---|---|---|---|
| | | $a$ | $b$ | $c$ | | | |
| Y123 | S950 | 3.82061 0.00004 | 3.88555 0.00005 | 11.68817 0.00010 | 1.68 | 3.0592 | 1.20 |
| | S980 | 3.82072 0.00004 | 3.88454 0.00005 | 11.68409 0.00010 | 1.66 | 3.0581 | 1.76 |
| Y134 | S950 | 3.80665 0.00010 | 3.88835 0.00006 | 15.26554 0.00038 | 2.12 | 4.0102 | 1.50 |
| | S980 | 3.80223 0.00017 | 3.88527 0.00017 | 15.25698 0.00040 | 2.16 | 4.0126 | 2.02 |
| Y145 | S950 | 3.80446 0.00015 | 3.86474 0.00013 | 19.37104 0.00023 | 1.57 | 5.0916 | 1.30 |
| | S980 | 3.80180 0.00014 | 3.86483 0.00005 | 19.38194 0.00017 | 1.64 | 5.0980 | 1.50 |
| Y156 | S950 | 3.80672 0.00012 | 3.88966 0.00008 | 22.90744 0.00063 | 2.16 | 6.0176 | 1.14 |
| | S980 | 3.80078 0.00011 | 3.89068 0.00006 | 22.94436 0.00063 | 2.33 | 6.0367 | 1.05 |

Table 4. This is the lattice parameters of non-superconducting compound.

| Compound | | Y211(Y$_2$BaCuO$_5$)  Pbnm  (Å) | | | BaCuO$_2$   Im-3m   (Å) | | | Ba$_2$Cu$_3$O$_6$   Pccm   (Å) | | |
|---|---|---|---|---|---|---|---|---|---|---|
| | | $a$ | $b$ | $c$ | $a$ | $b$ | $c$ | $a$ | $b$ | $c$ |
| Y123 | S950 | 7.19467 0.00183 | 12.16396 0.00409 | 5.64459 0.00223 | - | - | - | - | - | - |
| | S980 | 7.18778 0.00444 | 12.09504 0.00749 | 5.78043 0.00630 | - | - | - | - | - | - |
| Y134 | S950 | - | - | - | 18.44595 0.00032 | 18.44595 0.00032 | 18.44595 0.00032 | 13.03318 0.00014 | 20.64286 0.00029 | 11.40856 0.00017 |
| | S980 | - | - | - | 18.42390 0.00058 | 18.42390 0.00058 | 18.42390 0.00058 | 13.0268 0.00017 | 20.63403 0.00028 | 11.40945 0.00019 |
| Y145 | S950 | - | - | - | 18.23631 0.00021 | 18.23631 0.00021 | 18.23631 0.00021 | 12.99186 0.00015 | 20.56394 0.00012 | 11.37116 0.00014 |
| | S980 | - | - | - | 18.23979 0.00018 | 18.23979 0.00018 | 18.23979 0.00018 | 12.99265 0.00013 | 20.57803 0.00023 | 11.37412 0.00019 |
| Y156 | S950 | - | - | - | 18.30379 0.00024 | 18.30379 0.00024 | 18.30379 0.00024 | 13.04613 0.00020 | 20.62764 0.00029 | 11.40945 0.00018 |
| | S980 | - | - | - | 18.30325 0.00021 | 18.30325 0.00021 | 18.30325 0.00021 | 13.04909 0.00018 | 20.62446 0.00026 | 11.40088 0.00015 |

The thermal analysis measurements were studied and shown in Fig. 3. and Table 5. Our Y123 having the peritectic temperature in the same range of Y123 reported by Feng    et al.[22] as 1022 °C  and Xiang et al.[23] as 965 °C and 970 °C. The highest peritectic temperature was of Y123 and the lowest was of Y134 and Y156. Because the sintering temperature at 980 °C was higher than the onset melting point of our superconductors then some melting on the grain boundary was found by SEM.

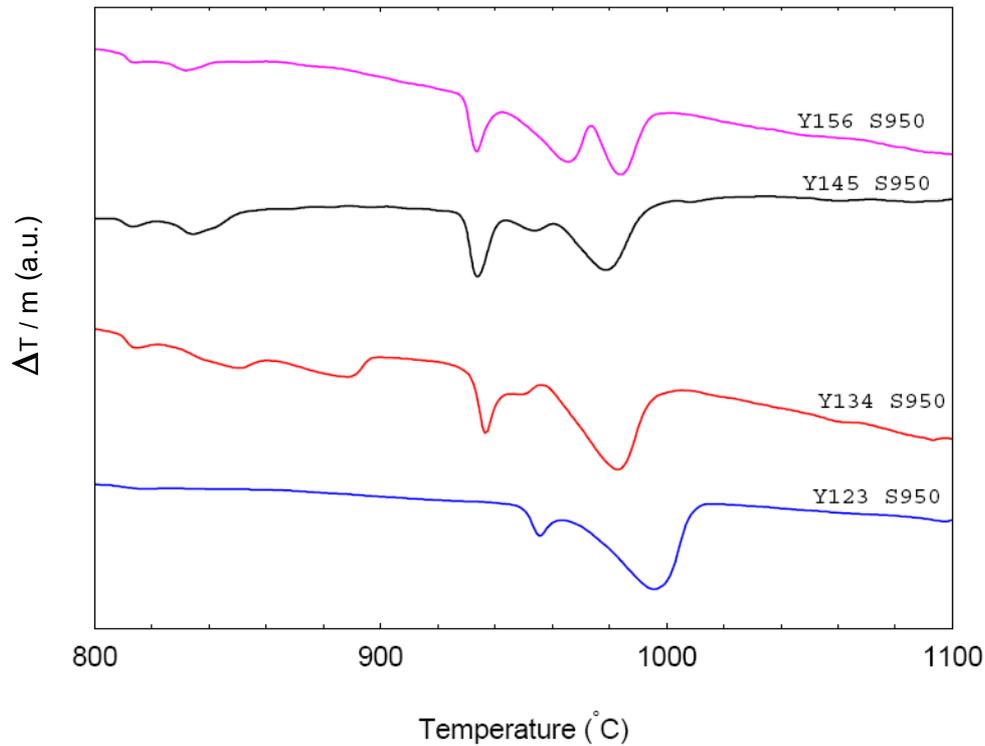

Fig. 3. This is the DTA curves of samples as a function of temperature for bulk Y123, Y145, Y134 and Y156.

Table 5. This is the melting point and onset melting point temperature of bulk Y123, Y134, Y145 and Y156.

| Samples | The onset melting point (°C) | The melting point (°C) |
|---------|------------------------------|------------------------|
| Y123 | 947 | 996 |
| Y134 | 930 | 983 |
| Y145 | 928 | 978 |
| Y156 | 928 | 983 |

The superconductivity properties of oxygen deficient perovskites, the crystal structure and oxygen content are closely related. A change in the oxygen content is induced a variation in properties of Y123. For a nearly optimally doped of $YBa_2Cu_3O_{7-y}$ ($y \approx 7$), a decrease in the oxygen content monotonically suppresses the superconductivity[24, 25]. The $YBa_2Cu_3O_{7-\delta}$, as $\delta$ varies in the range $1 \geq \delta \geq 0$ and the superconductivity was found in the range of $\delta$ from 0.2 to 0. For $0.5 < \delta \leq 1$ and $0 < \delta < 0.5$, the tetragonal-phase and the orthorhombic-phase of Y123 was found respectively. The phase transition from tetragonal to orthorhombic was occurred at $\delta$ = 0.5. For $\delta = 0$, the structure is completely orthorhombic-superconductive with $T_c$ above 90 K [26]. In the orthorhombic-phase while $b$-value increases, $a$- and $c$-values decrease with the increase of oxygen content were found. The $c$-value has been observed to change in the range from 11.8391 Å for $YBa_2Cu_3O_6$ (tetragonal) to 11.660 Å and for $YBa_2Cu_3O_7$ (orthorhombic) [26, 27].

The critical temperature has the relation to the amount of $Cu^{2+}$ and $Cu^{3+}$. The ratio of trivalent copper of Y123 ($YBa_2Cu_{2x}^{3+}Cu_{3-2x}^{2+}O_{6.5+x}$) with the $T_c^{onset}$ = 60 K ($x = 0.23$), 90 K ($x = 0.35$) and non-superconductor ($x \approx$ 0) were investigated by Choy et al.[28]. They also found that the ratio of $Cu^{3+}/Cu^{2+}$ depend on the annealing temperature and time. The higher $T_c$ is the higher of the $Cu^{3+}/Cu^{2+}$ becomes.

In this study, the standard Iodometric titration was used to determine the amount of $Cu^{2+}$ and $Cu^{3+}$ and oxygen content[29]. The oxygen content $O_y$ has been calculated by using the sum of the oxidation numbers of Y123 ($YBa_2Cu_3O_{7-x}$), Y134 ($YBa_3Cu_4O_{9-x}$), Y145 ($YBa_4Cu_5O_{11-x}$), Y156 ($YBa_5Cu_6O_{13-x}$) so they were $y = 7-x$, $9-x$, $11-x$, $13-x$, respectively. Here $x$ is the deficiency of samples. The $Cu^{3+}/Cu^{2+}$ ratio, Oxygen content and the deficiency of all samples were shown in Table 6. For all samples of our investigation, we find that the Y123 had shown the highest $Cu^{3+}/Cu^{2+}$, percentage of deficiency and critical temperature.

Table 6. This is the oxygen content and deficiency of samples.

| Compounds | | $Cu^{3+}/Cu^{2+}$ | Oxygen content($O_y$) | Deficiency (%) |
|-----------|------|-------------------|-----------------------|----------------|
| Y123 | S950 | 0.28 | 6.83 | 2.4 |
| | S980 | 0.29 | 6.84 | 2.3 |
| Y134 | S950 | 0.19 | 8.81 | 2.1 |
| | S980 | 0.19 | 8.81 | 2.1 |
| Y145 | S950 | 0.13 | 10.79 | 1.9 |
| | S980 | 0.12 | 10.77 | 2.1 |
| Y156 | S950 | 0.16 | 12.92 | 0.6 |
| | S980 | 0.14 | 12.86 | 1.1 |

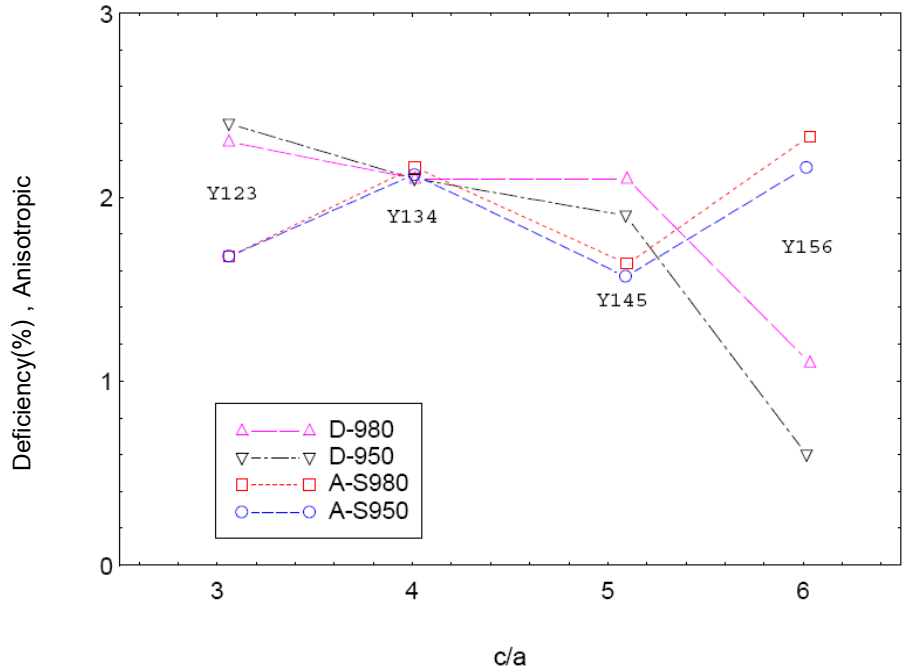

Fig. 4. This is the c/a ratio versus the D- Deficiency (%) and A- Anisotropic with varying the sintering temperature.

In Fig. 4, the effect of sintering temperature on the anisotropy and deficiency of our samples were shown. There was a little effect of changing sintering temperature on Y123 but more effect on Y156. The difference of anisotropy and deficiency of each samples increases as *c/a* increases. The Y156 has maximum anisotropy but minimum deficiency. Because the anisotropy of *YBaCuO* material was correspond to an orthorhombic distortion and asymmetric distribution of oxygen. Then Y156 has more orthorhombic distortion and asymmetric distribution of oxygen than the others.

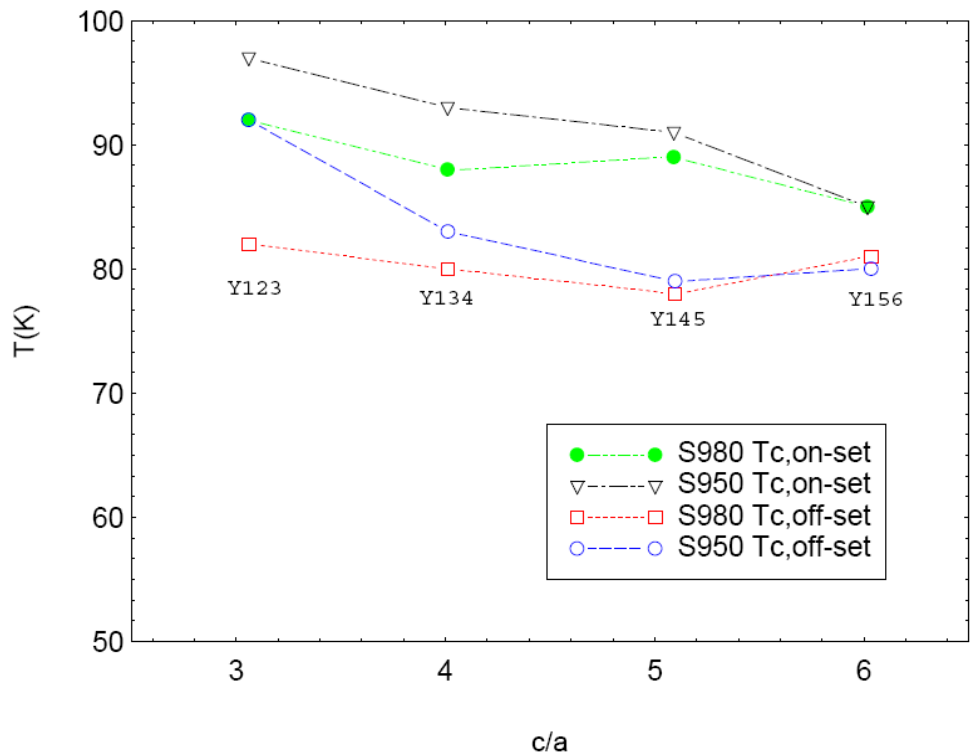

Fig. 5. This is the effect of sintering temperature on $T_c^{onset}$ and $T_c^{offset}$.

In Fig. 5, the effect of sintering temperature on $T_c^{onset}$ and $T_c^{offset}$ were shown. The Y123 had the highest critical temperature, $T_c^{onset} = 97$ K. The increasing of sintering temperature had dropped the critical temperature of these superconductors. However the increasing of sintering showed the effect on critical temperature indirectly that decrease oxygen content of Y134, Y145 and Y156 but not for Y123. The difference of $T_c^{onset}$ and $T_c^{offset}$ in each sample was decreases as $c/a$ increases.

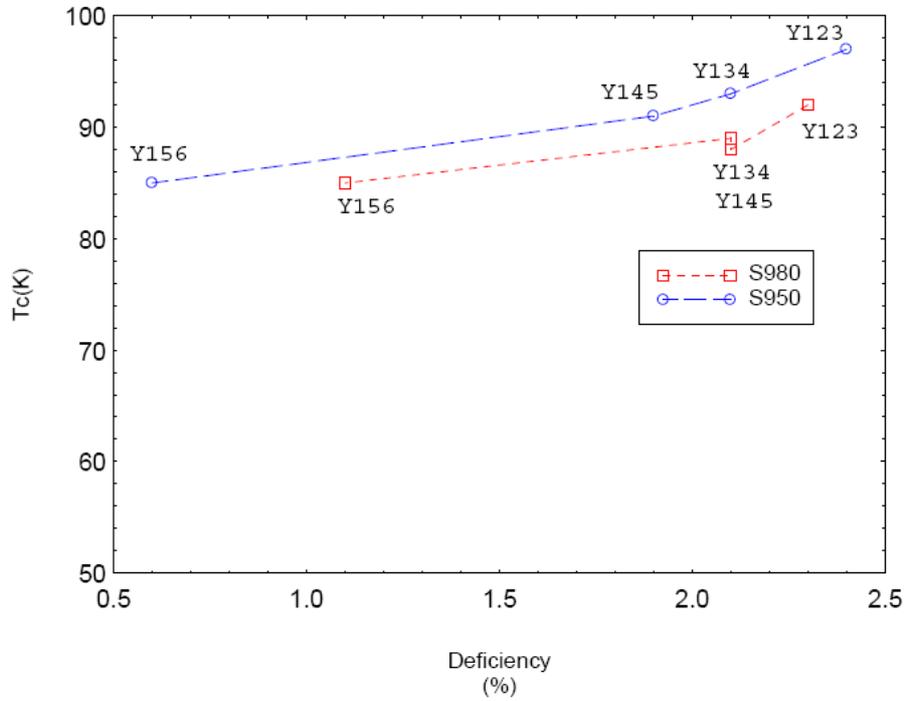

Fig. 6. This is the effect of sintering temperature and oxygen deficiency on the critical temperature onset of all samples.

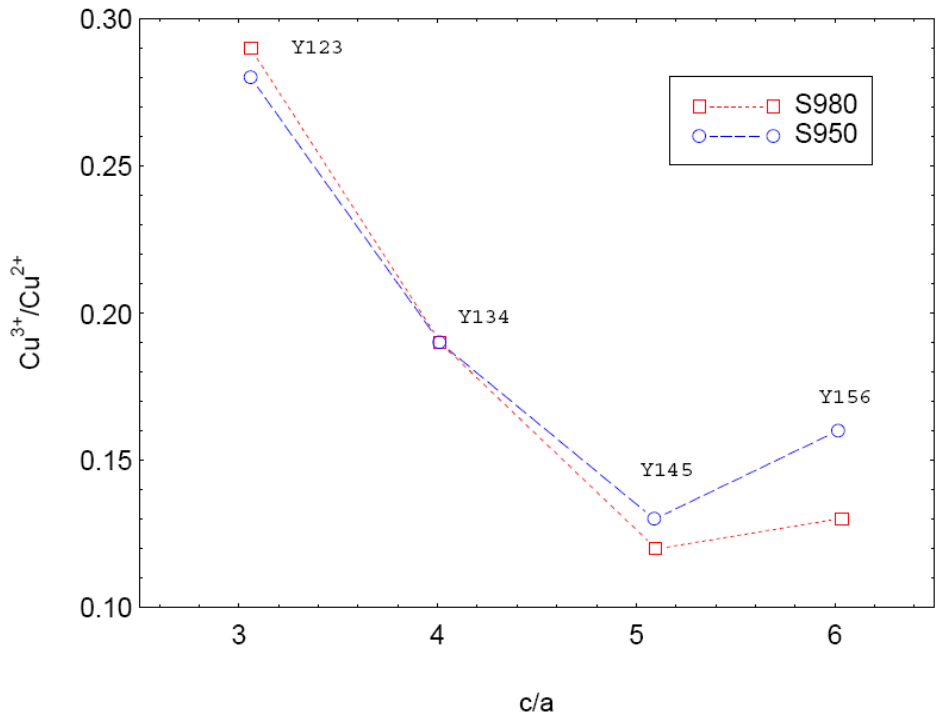

Fig. 7 Shown the ratio of $Cu^{3+}/Cu^{2+}$ versus $c/a$.

In Fig. 6., the effect of sintering temperature and oxygen deficiency on the critical temperature was shown. The relation of these parameter was almost be the linear dependence so we can conclude that the critical temperature depends on the deficiency and sintering temperature. The higher deficiency is, the higher critical temperature become. Then, the highest deficiency, Y123, has the highest critical temperature. The ratio of $Cu^{3+}/Cu^{2+}$ versus $c/a$ of all samples was shown in Fig. 7. We found that the ratio of $Cu^{3+}/Cu^{2+}$ depended on the sintering temperature. The Y123 has the maximum $Cu^{3+}/Cu^{2+}$ and the Y145 has minimum that agrees with the highest $T_c^{onset}$ of Y123 and the lowest $T_c^{offset}$ of Y145 (Fig. 5).

## 4. Conclusion

We synthesized the $YBa_mCu_{1+m}O_y$ superconductors; m = 2,3,4,5 that were Y123, Y134, Y145 and Y156 by solid state reaction. We found that the $T_c^{onset}$ of Y123, Y134, Y145, Y156 were at 97 K, 93 K, 91 K, 85 K and the ratio of $c/a$ were 3.0, 4.0, 5.0 and 6.0 and the ($Cu^{3+}/Cu^{2+}$ and Oxygen content) were (0.28, 6.83), (0.19, 8.81), (0.13, 10.79), (0.16, 12.92) respectively. The critical temperature depends on the deficiency. The higher deficiency is, the higher critical temperature become but the increasing of sintering temperature has reduced the critical temperature.

## 5. Acknowledgements

The author would like to thank the financial support of the Office of the Higher Education Commission, Pathumwan Institute of Technology, Srinakharinwirot University, and ThEP Center. And thank for Kiattipong Somsri, Nantawat Phomphuang, Prachkitti Mychareon, Kornkanit Kritcharoen, Pimpipa Butsingkorn and Pitanaree Ruttanaraksa for their help in data collection.